%% file: paine2019.tex
\documentclass[conference]{IEEEtran}
\IEEEoverridecommandlockouts
\usepackage{cite}
\usepackage{amsmath,amssymb,amsfonts}
\usepackage{algorithmic}
\usepackage{graphicx}
\usepackage[caption=false, font=footnotesize]{subfig}
\usepackage{textcomp}
\usepackage{xcolor}
\def\BibTeX{{\rm B\kern-.05em{\sc i\kern-.025em b}\kern-.08em
    T\kern-.1667em\lower.7ex\hbox{E}\kern-.125emX}}

\usepackage{xspace}
\usepackage{siunitx}
\usepackage{tikz}
\usepackage{circuitikz}
\usepackage{pgfplots}
\pgfplotsset{compat=1.14}
\usetikzlibrary{positioning}
\usetikzlibrary{patterns}
\usetikzlibrary{decorations.pathreplacing}
\usetikzlibrary{fadings}
\usetikzlibrary{circuits.ee.IEC}
\usetikzlibrary{fit}

\usepackage[skins]{tcolorbox}
\usepackage{hyperref}

\definecolor{scalebgcolor}{HTML}{000000} %

\newcommand{\vbatt}{ V\textsubscript{BATT}\xspace}

\begin{document}

\title{Evaluation of Low-Cost Thermal Laser Stimulation for Data Extraction and Key Readout}

\makeatletter
\newcommand{\linebreakand}{%
\end{@IEEEauthorhalign}
\hfill\mbox{}\par
\mbox{}\hfill\begin{@IEEEauthorhalign}
}
\makeatother

\author{\IEEEauthorblockN{Thilo Krachenfels}
\IEEEauthorblockA{\textit{Security in Telecommunications Group} \\
\textit{Technische Universität Berlin}\\
Berlin, Germany}
\and
\IEEEauthorblockN{Heiko Lohrke}
\IEEEauthorblockA{\textit{Security in Telecommunications Group} \\
\textit{Technische Universität Berlin}\\
Berlin, Germany}
\and
\IEEEauthorblockN{Jean-Pierre Seifert}
\IEEEauthorblockA{\textit{Security in Telecommunications Group} \\
\textit{Technische Universität Berlin}\\
Berlin, Germany}
\linebreakand
\IEEEauthorblockN{Enrico Dietz}
\IEEEauthorblockA{\textit{Institute of Optical Sensor Systems} \\
\textit{DLR}\\
Berlin, Germany
}
\and
\IEEEauthorblockN{Sven Frohmann}
\IEEEauthorblockA{\textit{Institute of Optical Sensor Systems} \\
\textit{DLR}\\
Berlin, Germany}
\and
\IEEEauthorblockN{Heinz-Wilhelm Hübers}
\IEEEauthorblockA{\textit{Institute of Optical Sensor Systems} \\
\textit{DLR}\\
Berlin, Germany}
}

\maketitle

\begin{abstract}
Recent attacks using thermal laser stimulation (TLS) have shown that it is possible to extract cryptographic keys from the battery-backed memory on state-of-the-art field-programmable gate arrays (FPGAs).
However, the professional failure analysis microscopes usually employed for these attacks cost in the order of 500k to 1M dollars.
In this work, we evaluate the use of a cheaper commercial laser fault injection station retrofitted with a suitable amplifier and light source to enable TLS. 
We demonstrate that TLS attacks are possible at a hardware cost of around 100k dollars.
This constitutes a reduction of the resources required by the attacker by a factor of at least five.
We showcase two actual attacks: data extraction from the SRAM memory of a low-power microcontroller and decryption key extraction from a $20$~\si{\nano\meter} technology FPGA device.
The strengths and weaknesses of our low-cost approach are then discussed in comparison to the conventional failure analysis equipment approach.
In general, this work demonstrates that TLS backside attacks are available at a much lower cost than previously expected.
\end{abstract}

\begin{IEEEkeywords}
IC Security, Optical Attacks, Thermal Laser Stimulation, FPGA Security
\end{IEEEkeywords}

\input{introduction.tex}
\input{background.tex}
\input{setup.tex}
\input{results.tex}
\input{discussion.tex}
\input{conclusion.tex}

\section*{Acknowledgment}
The authors would like to thank ALPhANOV for providing a Single Laser Microscope Station (S-LMS) for the setup.

\vspace{.5cm}
This is a pre-print of an article published in J Hardw Syst Secur. The final authenticated version is available online at: \url{https://doi.org/10.1007/s41635-019-00083-9}.

\bibliographystyle{IEEEtran}
\bibliography{IEEEabrv,literature}

\end{document}

%% file: introduction.tex
\section{Introduction}
Data extraction from integrated circuits (ICs) can pose a serious threat to the secrets stored within.
Extraction of cryptographic keys, sensitive data stored in memory, or device fingerprint information, as used in physically unclonable functions (PUFs), allows attackers to break security features.
Physical attacks, such as side-channel attacks, are one of the main approaches to extract data contained in embedded devices.

Thermal laser stimulation (TLS) is one such technique, which analyzes changes in the current consumption of the device in response to applied laser radiation.
In the past, it has been used to read out the content of static random-access memory (SRAM) and thus allows the characterization of SRAM PUFs \cite{nedospasov_invasive_2013}.
It was also applied to extract the key from the battery-backed random-access memory (BBRAM) contained within the decryption unit of a $20$~\si{\nano\meter} technology field-programmable gate array (FPGA) \cite{lohrke_key_2018}.
Furthermore, TLS can be considered as a suitable technique for the readout of microcontroller SRAM working memory \cite{kiyan_comparative_2018}.
Therefore, it is a powerful data extraction tool for an attacker on hardware level.

However, all previously mentioned experiments have been conducted using professional failure analysis (FA) equipment, more specifically a Hamamatsu Phemos-1000 laser scanning microscope (LSM).
Such a system typically costs around 500k to 1M dollars, and even when renting, costs for the development of a TLS attack are still in the range of thousands of dollars \cite{lohrke_key_2018}.
As a consequence, even though TLS is a powerful attack technique, the connected costs might discourage attackers from applying it.

Yet, it needs to be kept in mind that FA equipment usually offers a lot more features than an attacker might actually need, for instance, support for wafer handling and automated testing equipment, very fast acquisition times, and integration of other measurement techniques, such as photon emission.
In principle, however, all that is needed for a TLS attack is a way to move a laser spot over the device and simultaneously measure a current.
This raises the question, if attackers might be able to use simpler, more low-cost setups.
If so, the threat posed by TLS techniques would be larger than expected so far.
The main aim of this work is to determine if this is the case.

To evaluate this question, suitable alternatives to the usually employed FA systems need to be considered.
One such candidate are commercially available setups used for evaluation of laser fault injection (LFI) which are by a factor of around five to ten cheaper than FA LSMs.
Such systems usually feature a laser with focusing optics and some mechanical means to move the laser spot on the device under test (DUT), e.g., via motorized stages.
The only thing required to perform TLS with such a system would thus be a current preamplifier and a laser of suitable wavelength.
Hence, it seems plausible that such a setup could be modified to perform TLS attacks at a low cost.
However, it is unclear if the expected slower scanning speeds of motorized stages, as opposed to galvanometric mirrors usually used in FA solutions, might make attacks infeasible.
Besides that, a drift in electronics and the mechanical system as well as a lower scan resolution might hinder an attack.
An evaluation of the general possibility of developing such a setup thus seems to be beneficial.
Consequently, this paper will evaluate if such a setup is generally feasible and what advantages and disadvantages it would bring for a potential attacker.
This knowledge could then be used in the future to develop a more accurate TLS attacker model and thus better protected devices.

{\bfseries Our Contribution.}
In this work, we demonstrate the feasibility of a low-cost TLS attack setup by retrofitting a commercial LFI setup with a suitable laser, amplifier, and software.
For this, we only use commercially available components.
We then evaluate two previously published attack types on the setup.
The first one is the extraction of data from the SRAM of a microcontroller, as used for PUF characterization \cite{nedospasov_invasive_2013} and working memory data extraction \cite{kiyan_comparative_2018}.
For this attack, we showcase TLS scans of the whole memory area and also demonstrate that data can be extracted from the individual memory cells.
The second evaluated attack is the readout of the decryption key from the BBRAM of an FPGA, as presented in \cite{lohrke_key_2018}.
We demonstrate that even with a low-cost setup, extraction of the full 256-bit AES key from the device is possible.
Finally, we discuss and compare our results to the classical approach of using professional failure analysis equipment and highlight possible countermeasures.

%% file: background.tex
\section{Background}
\label{sec:background}

\subsection{Thermal Laser Stimulation (TLS)}
\label{sec:background:TLS}

\begin{figure}[tb]
	\centering
	\includegraphics[width=.75\linewidth]{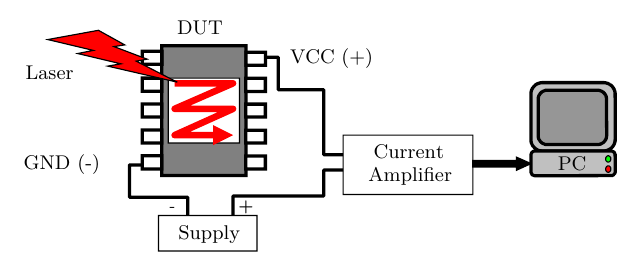}
	\caption{Scanning the laser over the DUT causes a change in current consumption due to thermal stimulation. Figure based on~\cite{beaudoin_implementing_2001}. \label{fig:background:tls_setup}}
\end{figure}

Techniques from failure analysis (FA) that use laser radiation to impact the device under test (DUT) are referred to as laser stimulation techniques.
Usually, the laser is scanned over the DUT while device parameters like the current consumption are monitored, grayscale-encoded and plotted over the scanning position, see Fig.~\ref{fig:background:tls_setup}.
The resulting response map shows areas where laser radiation causes changes in the current consumption of the DUT.
For thermal laser stimulation (TLS), the laser wavelength is chosen to have a photon energy smaller than the silicon bandgap, which consequently only causes local heating and no photocarrier generation.

When drain or source of a single metal-oxide semiconductor field-effect transistor (MOSFET) are thermally stimulated, effectively a voltage source between the corresponding metal contact and the channel is generated \cite{nedospasov_invasive_2013, boit_ultra_2013}.
This voltage source is also referred to as Seebeck generator, since it is caused by the Seebeck effect \cite{geballe_seebeck_1955}.
When the channel of the transistor is low-ohmic, this generator is connected between drain and source.
In contrast, when the channel is high-ohmic, one connection of the generator is floating and the generated voltage is ineffective.
The sign of the generated voltage depends on whether drain or source are stimulated and on the type of the MOSFET (n- or p-type) \cite{geballe_seebeck_1955}.

\begin{figure}[tb]
	\centering
	\includegraphics[width=\linewidth]{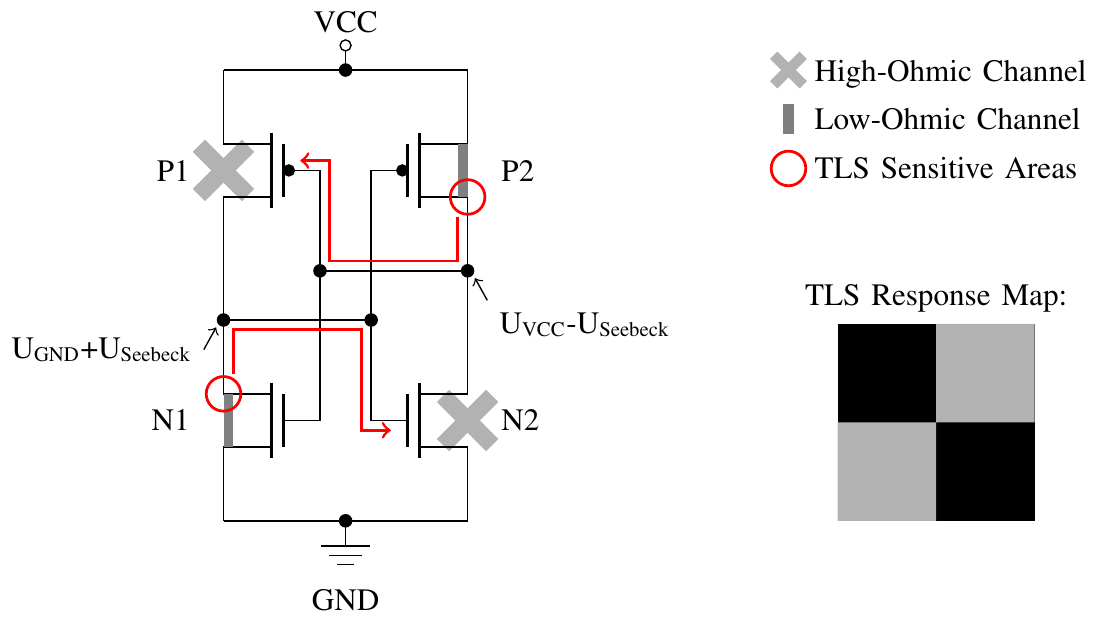}
	\caption{Memory cell under thermal stimulation. Figure based on~\cite{nedospasov_invasive_2013}. \label{fig:background:sram_cell}}
\end{figure}

A memory cell, as implemented in complementary metal-oxide semiconductor (CMOS) technology, basically consists of two cross-coupled inverters, see Fig.~\ref{fig:background:sram_cell}.
As can be seen, the circuit stays in one of two states because of the cross-coupling.
While being in a stable state, for ideal transistors there is no current flow between VCC and GND, since one of the transistors in each connection from VCC to GND is high-ohmic.
However, under stimulation, the Seebeck generator causes the creation of a voltage (U\textsubscript{Seebeck}), which is added to the existing voltage levels.
When assuming $0$~\si{\volt} as GND level and, for instance, the drain of transistor N1 is stimulated, effectively U\textsubscript{Seebeck} is applied to the gate of N2.
Consequently, the resistance of N2 decreases via exponential sub-threshold operation, which in turn results in an increased current flow between VCC and GND.
The same applies for transistor P1 when the drain of P2 is stimulated.
The change in current consumption can be expected to be in the nanoampere range \cite{nedospasov_invasive_2013}.
If a laser with a beam diameter approximately equal to the transistor size is scanned over the cell, a TLS response map as shown in Fig.~\ref{fig:background:sram_cell} can be expected.
Due to the increased current consumption, the sensitive transistors will be shown as brighter pixels. If the memory cell is in the inverted state, the other two transistors are sensitive.
The cell's state can thus be deduced from the TLS response map.

Note that due to the chosen laser wavelength only thermal stimulation occurs, which can increase the leakage current of the memory cell but cannot change its state.

\subsection{TLS for PUF Characterization and Data Extraction}
\label{sec:background:TLS_SRAM}

The extraction of data stored in SRAM on microcontrollers can pose a threat to secrets stored within.
For instance, the authors of \cite{nedospasov_invasive_2013} show that the extraction of data from SRAM memory on microcontrollers down to the $180$~\si{\nano\meter} technology node is possible.
More specifically, they demonstrate the characterization of a proof-of-concept SRAM-based PUF implementation on a microcontroller using TLS on professional FA equipment.
Similarly, the authors of \cite{kiyan_comparative_2018} show the potential to read out the whole working memory on a $180$~\si{\nano\meter} technology microcontroller using TLS.
It should be noted that for both attacks it was necessary to put the DUTs into a low-power mode, to reduce the noise of the system.

Such attacks on SRAM memory of microcontrollers are hereafter referred to as SRAM data extraction attacks.

\subsection{TLS for Decryption Key Extraction}
\label{sec:background:TLS_BBRAM}

The authors of \cite{lohrke_key_2018} demonstrate that the battery-backed random access memory (BBRAM) on a $20$~\si{\nano\meter} technology field-programmable gate array (FPGA) can be read out with TLS using professional FA equipment. The BBRAM stores a 256-bit key used for bitstream decryption.
To retain the key while the FPGA is powered off, the BBRAM is powered by a coin-cell battery.
During the attack, the TLS signal is acquired by measuring the current consumption on this battery line.
Since the BBRAM is the only circuit powered by the battery, the noise on this battery line is very low.
It should be noted that the attack was successful because the memory cell size is approximately $2.8 \times 3.1$~\si{\micro\meter}, which is about 10 times larger than expected  minimum size on a $20$~\si{\nano\meter} technology device \cite{lohrke_key_2018}.
This can be explained by reliability, leakage, and low current consumption considerations.

For their attack approach, the authors assume that the BBRAM is located close to the configuration logic. They consult the documentation to get an estimate of its location on the chip. By conducting a TLS scan over the candidate area, they can find the BBRAM. Afterward, they prove a data dependency in the measurements and create a mapping from memory cell locations to logical bits. Finally, they show that a key stored in the BBRAM can be extracted using TLS in a manual or automated fashion within minutes.

Although the BBRAM is typically only battery-backed SRAM, this attack type is hereafter referred to as BBRAM key readout attack.

%% file: setup.tex
\section{Setup}

\subsection{Laser Stimulation Setup}
As core of our setup we use an ALPhANOV Single Laser Microscope Station (S-LMS), which was designed for laser fault injection purposes \cite{alphanov_optical_2019}.
It is a microscope-based setup that allows the injection of different laser sources.
In our case we use a $1424$~\si{\nano\meter} laser diode capable of delivering more than 300 mW in continuous waveform (CW) mode.
The laser power can be controlled via PC software.
The laser is focused through objectives, which are mounted on a manual turret, into the IC backside.
For thermal stimulation we use a 50x/0.65NA objective with silicon thickness correction, for optical images we additionally use 20x/0.5NA and 2.5x/0.1NA objectives.
The whole microscope is mounted on XYZ motorized stages, which allow movements with a resolution of $50$~\si{\nano\meter}.
The stages are controlled via PC software or a joystick.
The S-LMS is also equipped with infrared (IR) lighting and a short-wave infrared (SWIR) camera.
This allows the user to monitor the laser spot position and perform optical navigation.

For measuring the current consumption during stimulation, we use a Stanford Research Systems “SR570” current preamplifier, which has a bias voltage feature.
The preamplifier outputs  a voltage proportional to the current which is digitized using a National Instruments "PCI-6259" card. 

\begin{figure}[tb]
	\centering
	\includegraphics[width=\linewidth]{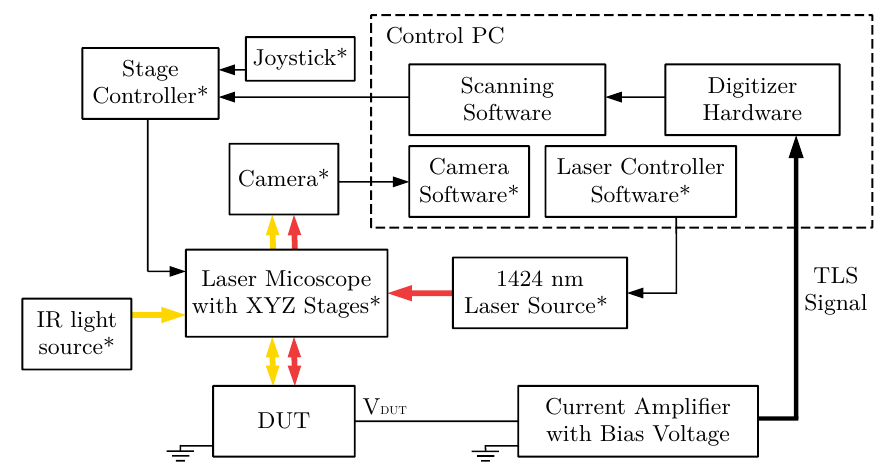}
	\caption{Block diagram of the setup. Components marked 
	with * are part of the S-LMS \cite{alphanov_optical_2019}. \label{fig:setup:schematic}}
\end{figure}

To realize the scanning functionality and TLS response map creation, we developed a scanning software in the ``LabView'' programming environment from National Instruments.
In this software, the scanning parameters, such as step size, step resolution, scanning speed and number of samples per pixel can be entered.
The stage, and thus also the laser spot, is then moved continuously over the DUT while the preamplifier output is sampled.
From this data, a TLS response map is created, in which  higher current consumption of the DUT corresponds to brighter pixels.
The whole setup is visualized in Fig.~\ref{fig:setup:schematic}.
Note that all results shown in this work have been achieved with this setup.

\subsection{Devices Under Test (DUTs)}
\label{sec:setup:DUTs}

\subsubsection{DUT for SRAM Data Extraction}
\label{sec:setup:DUTs_MSP}

As mentioned in Sect.~\ref{sec:background:TLS_SRAM}, reading out SRAM via TLS has been demonstrated down to the $180$~\si{\nano\meter} technology node.
Thus, a $180$~\si{\nano\meter} Texas Instruments MSP430F5131 microcontroller is used in our experiments.
It is equipped with 1~KB of SRAM with a cell size of approximately $2.5 \times 1.9$~\si{\micro\meter} \cite{kiyan_comparative_2018}.
For access to the silicon, the backside packaging material and the metal chip carrier were removed.

During the experiments, VCC of the DUT is supplied with $2.6$~\si{\volt} via an auxiliary power supply.
The core, which contains the SRAM, is supplied via the internally generated VCORE voltage, which is also available externally at a pin.
To this pin we connect the SR570 current preamplifier and set the bias voltage to $2.1$~\si{\volt}, which is slightly above the VCORE voltage of $1.9$~\si{\volt}.
In this way, a significant amount of the core voltage is supplied by the SR570.

A JTAG debugging interface is connected to the device.
This allows to directly write arbitrary data into the SRAM.
For noise reduction on the VCORE net, the DUT is send
 to low-power mode 4 (LPM4) during the TLS scan.

For all experiments on the MSP430, the current amplification of the SR570 was set to $1$~\si{\nano\ampere\per\volt} and the input offset to $500$~\si{\nano\ampere}.
The laser current was set to $600$~\si{\milli\ampere}, which corresponds to a total power of about $43$~\si{\milli\watt} for the 50x lens.
The silicon thickness correction of the objective was set to $350$~\si{\micro\meter}.

\subsubsection{DUT for BBRAM Key Readout}
\label{sec:setup:DUTs_Ultrascale}

The target platform for bitstream key extraction is a Xilinx Ultrascale FPGA development board from AVNET (model AES-KU040-DB-G).
It contains a Xilinx Ultrascale XCKU040-1FBVA676 FPGA manufactured with 20 nm technology in a flip-chip ball grid array (BGA) package.
Due to the flip-chip package, direct access to the silicon is available and no preparation is necessary.
The thickness of the substrate is about $750$~\si{\micro\meter} \cite{lohrke_key_2018}.

The 256-bit key used for bitstream decryption can be stored in a battery-backed RAM (BBRAM) which is programmed via a JTAG interface \cite{xilinx_using_2017}.
While the device is powered off, the BBRAM is supplied by a battery via the \vbatt line.
To measure the current consumption of the BBRAM during TLS, we soldered cables to the battery connector and connected them to the input of the SR570 current amplifier.
During key programming, the board is powered via its external supply.
During TLS experiments, however, the board is powered off and the \vbatt voltage is supplied via the bias voltage feature of the SR570.

For all experiments on the Ultrascale FPGA, the current amplification of the SR570 was set to $2$~\si{\nano\ampere\per\volt} with no input offset.
The laser current was set to $500$~\si{\milli\ampere}, which corresponds for the 50x lens to a total power of about $26$~\si{\milli\watt}.
The silicon thickness correction of the objective was set to $750$~\si{\micro\meter}.

%% file: results.tex
\section{Measurement Results}

\subsection{SRAM Data Extraction}
\label{sec:results:SRAM}

\subsubsection{SRAM Overview}
\begin{figure}[tb]
	\centering
	\includegraphics[width=.7\linewidth]{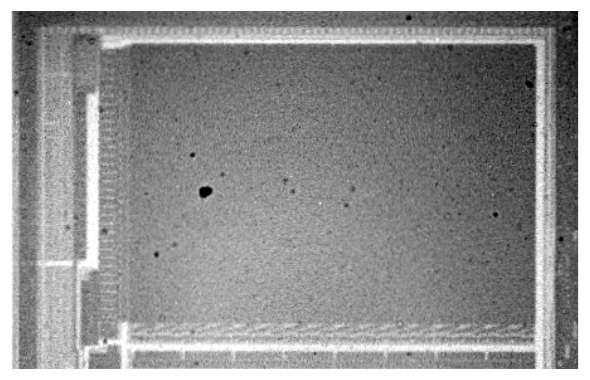}
	\caption{Optical image (20x lens) of the SRAM block. \label{fig:results:msp_overview_camera}}
\end{figure}

\begin{figure}[tb]
	\centering
	\includegraphics[width=.8\linewidth]{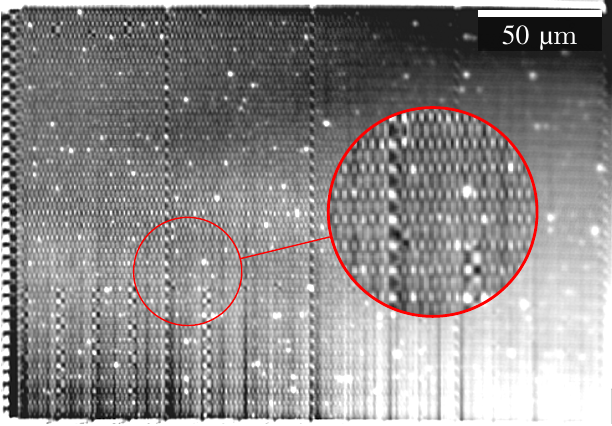}
	\caption{TLS overview scan of the SRAM. The 
scan direction is bottom-to-top (fast axis) and 
then left-to-right (slow axis). \label{fig:results:msp_overview_TLS}}
\end{figure}

To localize the  SRAM optically, the camera of the setup was used, see Fig.~\ref{fig:results:msp_overview_camera}.
After zeroizing the whole SRAM via JTAG, the device is sent to low-power mode by code run from flash.
A TLS scan with $0.5$~\si{\micro\meter} scan step size was then acquired,  see Fig.~\ref{fig:results:msp_overview_TLS}.
It can be seen that most of the memory shows a regular structure, except for some irregular vertical strips, mainly in the bottom left quadrant.
Closer investigation revealed that this is data placed in SRAM by the code which enters the low-power mode.
This already demonstrates that data dependencies can be observed.
The SRAM seems to be sub-divided into four blocks with a small offset of about one cell width in between, as already discovered in \cite{kiyan_comparative_2018}.
In addition, some cells seem to be more sensitive to TLS, as can be seen by some irregular bright spots.
This can be explained by manufacturing variability.
The TLS response becomes increasingly blurry in the right half of the scan, which can be explained by thermal and mechanical drift of the DUT due to the long scan duration of 43~\si{\minute}.
This could be avoided by scanning smaller areas and refocusing for each measurement.

\subsubsection{Extraction of Single Bits}

\begin{figure}[tb]
	\centering
	\includegraphics[width=.75\linewidth]{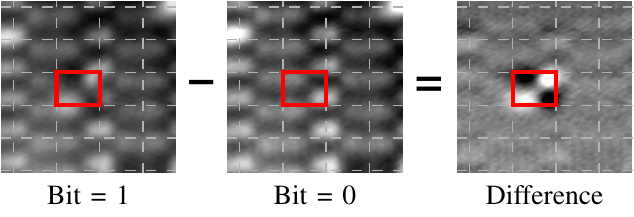}
	\caption{Data dependency of the measurements for a single SRAM bit which was first set to \texttt{1} and then to \texttt{0}, while all other bits in the area are set to \texttt{0}. The subtraction of both response maps reveals the change more clearly. \label{fig:results:msp430:singleBit}}
\end{figure}

To demonstrate the extraction of single bits, we compare measurements of a small area of the SRAM, see Fig.~\ref{fig:results:msp430:singleBit}.
For the first measurement, the centered bit (framed in red) is set to \texttt{1}, while all other bits are  \texttt{0}.
For the second measurement, all bits, including the highlighted one in the center, are \texttt{0}.
A pattern similar to the response map in Fig.~\ref{fig:background:sram_cell} can be observed.
For bit value \texttt{1}, bottom left and top right of the cell are the most sensitive spots.
In contrast, for bit value \texttt{0}, the sensitive spots are in the other corners of the cell.
The subtraction of both response maps reveals the change more clearly.

These results show that the resolution of our setup is sufficient for extracting data from arbitrary SRAM cells on the MSP430 device.
Consequently, an SRAM PUF implemented with a similar feature size could be characterized with this setup.
If the memory layout would be reverse-engineered using TLS,  the full working memory of the MSP430 could be read out as well.

\subsection{BBRAM Key Readout}

\subsubsection{Localization and Optical Overview}

\begin{figure}[tb]
	\centering
	\subfloat[BBRAM activated.\label{fig:results:ultrascale:localization_on}]{
		\includegraphics[width=.45\linewidth]{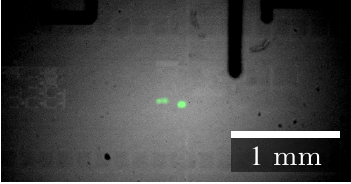}
	}
	\subfloat[BBRAM deactivated.\label{fig:results:ultrascale:localization_off}]{
	\includegraphics[width=.45\linewidth]{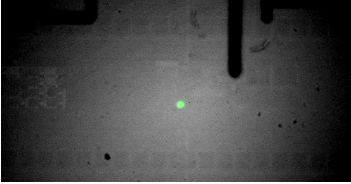}
	}
	\caption{Localization scan with activated and deactivated BBRAM. The TLS signal is superimposed on an optical image (2.5x lens). \label{fig:results:ultrascale:localization}}
\end{figure}

\begin{figure}[tb]
	\centering
	\includegraphics[width=.7\linewidth]{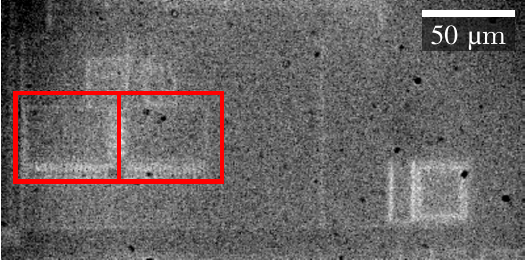}
	\caption{Optical image (20x lens) of the area sensitive to TLS with BBRAM candidate framed red. \label{fig:results:ultrascale:overview20x}}
\end{figure}

In the attack scenario of \cite{lohrke_key_2018}, the BBRAM has first to be localized inside the configuration area.
For this, we performed a scan of that area with a pixel size of $5$~\si{\micro\meter} and a stage speed of $2$~\si{\milli\meter\per\second} in about $5$~\si{\minute}.
The response map, see Fig.~\ref{fig:results:ultrascale:localization_on}, reveals two sensitive areas when the BBRAM is activated.
If the BBRAM is deactivated, only one sensitive area remains, see Fig.~\ref{fig:results:ultrascale:localization_off}.
Fig.~\ref{fig:results:ultrascale:overview20x} shows an optical image of the area where TLS sensitivity occurred.
The two highlighted block-like structures on the left correspond to the area where the TLS signal was dependent on BBRAM activation.
These are the BBRAM block candidates, which are already known from \cite{lohrke_key_2018}.
The structure on the right-hand side was always sensitive and thus can be disregarded.
The results show that the BBRAM can be localized using our setup.
In the next step, the detailed TLS response map has to be analyzed.

\subsubsection{TLS Overview Scan}

\begin{figure}[tb]
	\centering
	\subfloat[Random key \label{fig:results:ultrascale:randKey}]{
		\begin{tikzpicture}
				\begin{scope}
				\node[inner sep=0] (img) at (0,0) {\includegraphics[width=.47\linewidth, angle=180]{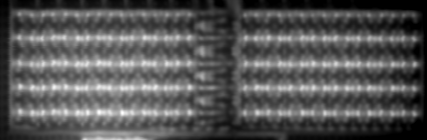}};
				\end{scope}
				\end{tikzpicture}
	}
	\subfloat[All-zeroes key \label{fig:results:ultrascale:zeroKey}]{
		\begin{tikzpicture}
				\begin{scope}
				\node[inner sep=0] (img) at (0,0) {\includegraphics[width=.47\linewidth, angle=180]{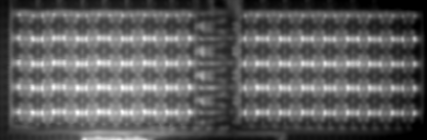}};
				\end{scope}
				\end{tikzpicture}
	}
	\caption{TLS response maps of the whole BBRAM programmed with two different keys. \label{fig:results:ultrascale:randZeroKey}}
\end{figure}

To observe data dependencies in the TLS response of the BBRAM, we first programmed a random key and an all-zeroes key and acquired TLS response maps, see Fig.~\ref{fig:results:ultrascale:randKey} and \ref{fig:results:ultrascale:zeroKey}.
It can be seen that different keys lead to different patterns in the response map.

To further investigate the key dependency of the TLS response, a single memory cell can be examined.

\subsubsection{Extraction of a Single Bit}

\begin{figure}[tb]
	\centering
	\includegraphics[width=.75\linewidth]{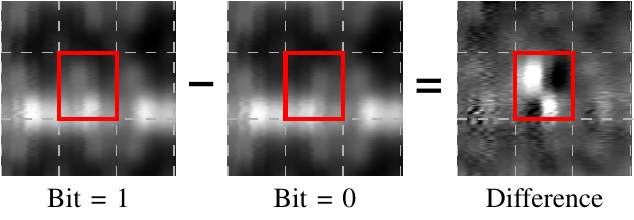}
	\caption{Data dependency of the measurements for a single BBRAM bit which was first set to \texttt{1} and then to \texttt{0}, while all other bits in the area were \texttt{0}. The subtraction of both response maps  highlights the change more clearly. \label{fig:results:ultrascale:singleBit}}
\end{figure}

To identify a single bit in the TLS response map, we scanned a small area of the BBRAM with high resolution (pixel size $50$~\si{\nano\meter}, stage speed $50$~\si{\micro\meter\per\second}) with different bit values for one memory cell, see Fig.~\ref{fig:results:ultrascale:singleBit}.
While the sensitive spots for bit value \texttt{1} are on the top left and bottom right, the spots for value \texttt{0} are on the top right and bottom left of the cell, cf. Fig.~\ref{fig:background:sram_cell}. 
The subtraction of the two response maps clearly shows that the state and thus the bit value of the centered BBRAM cell differs in the two measurements.
Hence, this experiment proves that the optical resolution of our setup is sufficient for extracting the bit value stored in one BBRAM cell.
The observed cell size is about $3.2 \times 2.8$~\si{\micro\meter}.

\subsubsection{Key Extraction}

\begin{figure}[tb]
	\centering
	\includegraphics[width=.9\linewidth]{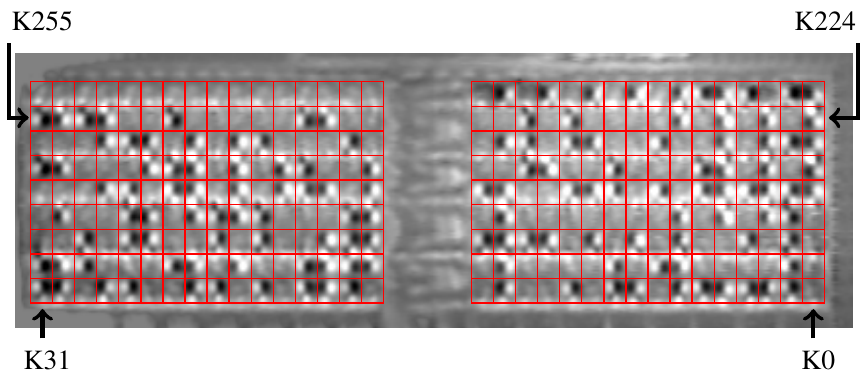}
	\caption{Subtraction of the TLS response maps of the random key and the all-zeroes key. The existence of black and white patterns in one cell corresponds to bit value \texttt{1}. Bit position K255 corresponds to the most significant bit of the key. The key programmed into the BBRAM is: \texttt{0xf2\-0c\-28\-55\-1d\-62\-6c\-97\-c7\-59\-32\-35\-1b\-5d\-ce\-bf\-4d\-e3\-40\-56\-2c\-a7\-f5\-4a\-e3\-4f\-42\-c2\-d9\-ae\-4b\-7e}. \label{fig:results:ultrascale:diffKeys}}
\end{figure}

Since we have already shown that data extraction of single bits from the BBRAM is possible with our setup and the mapping from physical to logical bit positions is known from \cite{lohrke_key_2018}, now a complete key can be extracted.
For this, we subtract the response map of the all-zeroes key (Fig.~\ref{fig:results:ultrascale:zeroKey}) from the response map of the random key (Fig.~\ref{fig:results:ultrascale:randKey}).
On the difference image, see Fig.~\ref{fig:results:ultrascale:diffKeys}, areas with large black and white spots correspond to bit value \texttt{1}, the others to value \texttt{0}.
By adding a grid to optically show the SRAM cell size and position, the key bits can be easily extracted manually. Note that the top row is used to store security-relevant information, such as a configuration counter and error-detection bits \cite{lohrke_key_2018}.
The scan of the whole BBRAM for a pixel size of $250$~\si{\nano\meter} and a stage speed of $50$~\si{\micro\meter\per\second} takes about $7$~\si{\minute}.

Given the above, the bitstream decryption key can be extracted from the BBRAM using our setup within minutes.
This proves that the complete attack on the FPGA bitstream decryption key can be conducted with a much cheaper setup than previously expected.

%% file: discussion.tex
\section{Discussion}

\subsection{Low-Cost vs. FA Setup}

\subsubsection{Acquisition Time}
The experiments have shown that the time needed for acquisition is substantially longer when using the low-cost setup.
This is due to the fact that for an FA laser scanning microscope (LSM) only small and light galvanometric mirrors are moved to scan the beam.
In our case, though, the optical setup is moved on mechanical stages and the connected inertia poses a limit on the maximum scan speed.
To give some exemplary numbers, for an FA LSM an acquisition time of $1.2$ minutes can be expected for BBRAM key extraction \cite{lohrke_key_2018}.
With our setup, $7$ minutes were needed.
This is an increase by a factor of $5.8$, but makes measurements due to the generally short duration still unproblematic.
For TLS SRAM data extraction, experiments performed by the authors have resulted in $4.8$ minutes of acquisition time on an FA LSM.
Using the low-cost setup, a complete SRAM scan on the MSP430 takes $43$ minutes.
This is an increase by a factor of $9$.
For such a long acquisition time, negative effects such as sample drift will lead to complications.
This could already be observed in Fig.~\ref{fig:results:msp_overview_TLS} (Sect.~\ref{sec:results:SRAM}).
It can thus be seen that with the approach demonstrated in this paper, attackers will have to trade cost for time.
Additionally, procedures such as refocusing might be needed to prevent negative side effects, although these are relatively easy to implement. 

For a low-cost approach, mechanical stages seem to be the obvious choice, since they are available in virtually any laboratory and microscope setup, as also in the used fault injection setup.
However, it should be noted that galvanometric scan mirrors can be acquired at comparable prices to mechanical stages.
Yet, the optical setup is more demanding, especially the requirements on the objective rise, since the field of view has to be sufficiently large.

\subsubsection{Resolution}
In terms of optical resolution, FA LSMs and the low-cost approach are virtually identical.
This is due to the fact that the optical resolution is mainly determined by the wavelength and the numerical aperture of the objective lens.
Using the same lens and wavelength should thus yield the same resolution in both setups.
For the $50$x lens and laser used in our setup, an optical resolution of about $1$~\si{\micro\meter} can be expected.

For scan step resolution, the situation is different.
The angular stepping resolution of an LSM's scanning mirror is translated by the objective lens into a spatial scanning resolution.
This means that the scanning resolution can be increased by using larger magnification lenses.
In contrast, the scan step resolution for the low-cost setup is simply the resolution of the stage, $50$~\si{\nano\meter} in our case.
Both the LSM's and the low-cost setup's scan resolution are significantly lower than the optical resolution and can be expected to not be a limiting factor.

In general it can be said that our setup is not better or worse compared to an LSM in terms of resolution.
However, it should be noted that FA LSMs can be equipped with a solid immersion lens (SIL), which can increase the resolution by a factor of around $4.3$ in case of the Hamamatsu Phemos system \cite{hamamatsu_nanolens_2015}.

\subsubsection{Cost}
Our setup  only consists of commercially available components.
The core of the setup is a ``S-LMS'' station by ALPhANOV.
In a configuration suitable for retrofitting TLS, the system costs about 102k~USD, including a $1.4$~\si{\micro\meter} laser.
Additionally, the SR570 current preamplifier for 2,595~USD \cite{sr570} and the NI-6259 digitizer card for 1,940~USD \cite{national_instruments_ni_2019} have to be acquired.
Including the control PC and a LabView license, the price of the complete setup is expected to be below 110k~USD.

Compared to a Phemos-1000 Failure analysis setup with a price between 500k and 1M USD, our setup is five to ten times cheaper.
Furthermore, the setup can in principle be set up on a single desk and purchasing of the equipment is expected to require less effort.

\subsection{Attack Feasibility and Limitations}
The feasibility of TLS attacks on SRAM in general depends on the spatial distance between the sensitive transistors.
Consequently, the limiting factor is the laser spot size.
The minimum possible spot diameter is about $1$~\si{\micro\meter} without SIL and $235$~\si{\nano\meter} with a recent SIL, which corresponds to cell dimensions of $2$~\si{\micro\meter} and $470$~\si{\nano\meter}, respectively. 
Thus, the attack is expected to work at least down to these cell sizes.
Furthermore, it can be expected that with post processing of the TLS signal, for instance, by deconvolution, an additional resolution enhancement by a factor of two is possible.
However, to the best of our knowledge, the actual SRAM cell size limit for TLS attacks is unknown and should be subject of future research.
It should also be noted that with the switch to new technologies, like FinFET, changes in the behavior of the stimulated cells might occur.

Next to the cell size, the attack is also limited by the amount of noise present in the TLS signal.
Specifically, the leakage current of the transistor affected from stimulation should be higher than the fluctuations in the overall current consumption.
In our experiments, this was fulfilled by the low noise on the battery line of the FPGA, and by sending the microcontroller to low-power mode.

Readers interested in more details regarding the attack feasibility are directed to \cite{lohrke_key_2018}.

\subsection{Countermeasures}
The possible countermeasures against TLS attacks from the chip backside can be divided into two categories.
On the one hand, techniques could be applied to obstruct the access to the chip or the measurement signal, and on the other hand, active attack detection mechanisms could be employed.

An approach for the former class could be to reduce the resolution of the laser beam by scrambling the incoming light from the chip backside, and thus increasing the beam diameter within the silicon.
In \cite{shen_nanopyramid_2018} this approach is applied against optical contactless probing.
The authors introduce the usage of nanopyramid structures, which scramble the reflected light.
However, adding the nanopyramids between chip backside and the transistors is only possible for bonding-based SOI devices.
Furthermore, the countermeasure was not tested with respect to thermal stimulation.
Yet, it might be an interesting approach for further research.

Another approach for the first category of countermeasures could address the destruction of the data dependency in the TLS signal.
Since the attack relies on low noise in the current consumption of the target device, noise injection can be used for this purpose.
In \cite{das_high_2017} a noise source has been successfully designed and integrated to protect an encryption core from power analysis attacks.
This shows that on-chip noise-based mitigation techniques can work.
Against TLS attacks on SRAM, a proof of concept countermeasure was presented in \cite{lohrke_key_2018}.
By injecting noise on the battery line of a BBRAM key storage, TLS data extraction can be made much harder or possibly even unfeasible.
The authors show that this can be an effective mitigation technique, even with negligibly lower battery life time.

A more thorough approach is the protection of the chip backside by employing an opaque coating layer to obstruct optical access completely.
However, a solely passive layer could be easily removed by polishing.
As evaluated in \cite{amini_assessment_2018}, the integrity of the coating layer can be assured by in-silicon light emitters and sensors. This combines an obstruction approach (first category) with an active detection countermeasure (second category).
Yet, due to the high power consumption of the photo sensors, this protection scheme can not protect devices with a very restricted power profile, such as the BBRAM key storage.

To actively detect the temperature changes induced by the laser radiation, temperature sensors could be useful \cite{lohrke_key_2018}. However, it is questionable whether the very small temperature changes with a power of less than $50$~\si{\milli\watt} can be detected with a small amount of false positives. Furthermore, the current consumption of temperature sensitive circuits, such as ring oscillators, is typically high \cite{tajik_pufmon_2017}. Hence, for power constrained devices like the BBRAM, this does not seem to be a feasible solution.

%% file: conclusion.tex
\section{Conclusion}
In this work, we have shown that constructing a low-cost setup for TLS is indeed feasible.
By retrofitting an LFI setup with the necessary equipment, we have demonstrated a solution for TLS five to ten times cheaper than traditional FA equipment.
Although with slower signal acquisition, we were still able to show that two state-of-the-art attacks, specifically against SRAM on a microcontroller and BBRAM on an FPGA, are possible in reasonable time.
Consequently, the attacker model must be rethought and adapted to better reflect the lower-than-expected hurdle for an attacker to apply TLS.
Therefore, better protection mechanisms against attacks from the chip backside will have to be deployed.